\documentclass[preprints,article,accept,moreauthors,pdftex]{Definitions/mdpi} 

\usepackage{amssymb}
\usepackage{amsmath}

\usepackage{color}
\usepackage{soul}
\usepackage{siunitx}
\DeclareSIUnit{\ppm}{ppm}
\DeclareSIUnit{\dBm}{dBm}

\usepackage{mhchem}
\usepackage{upgreek}
\usepackage{mathtools}

\usepackage{stackengine}

\newdimen\slantmathcorr
\def\undersl#1{
\setbox0=\hbox{$#1$}
\slantmathcorr=\wd0
\underline{\hbox to 0.65\wd0{%
\vphantom{\hbox{$#1$}}}}
\hskip-0.77\wd0\hbox{$#1$}
}

\def\underslv#1{
\setbox0=\hbox{$#1$}
\slantmathcorr=\wd0
\underline{\hbox to 0.65\wd0{%
\vphantom{\hbox{$#1$}}}}
\hskip-0.80\wd0\hbox{$#1$}
}

\firstpage{1} 
\pubvolume{xx}
\issuenum{1}
\articlenumber{5}
\pubyear{2025}
\copyrightyear{2025}
\history{Received: October 7, 2026}

\Title{On the Electrodynamics of a Non-Contact Inductive Sensor for
High-Accuracy Tip Deflection Measurements of Piezoceramic Bending Actuators
and
its Application for Energy Harvesting Purposes}

\Author{Rüdiger G. Ballas $^{1}$\orcidA{}}

\AuthorNames{Rüdiger G. Ballas}

\address{%
$^{1}$ \quad Department of Engineering Sciences, Wilhelm Büchner Hochschule,
64925, Darmstadt, Germany}

\corres{Correspondence: ruediger.ballas@wb-fernstudium.de}

\abstract{Piezoelectric bending actuators are suitable for a wide range of
applications that require deflections in range of hundreds of microns. A
variety of applications can be found in almost all fields of electrical
engineering, mechanical engineering, acoustics, automation, automotive, health
care and in countless further application for industry and daily life. The
development of high-efficient and low-cost piezoceramic materials optimized for
technical applications allow a variety of technical solutions for piezoceramic
actuators. The high-effective mass systems provide maximum gain at only
$\SI{1}{\kilo\volt/\milli\metre}$ resulting in a material strain of
approx.
$\SI{2}{\ppm}$. Thus, high-effective piezoceramics show very large
piezoelectric
coefficients $d_{iq}$ representing the ratio between strain and electric field.
However, inherent hysteresis, drift and creep characteristics are a
disadvantage of high-effective piezoceramic materials. Therefore, piezoceramic
bending actuators consisting of high-efficient piezoceramics are only
suitable
for a limited extent, as far as high-accuracy positioning is concerned. For
applications, that require precise switching and exact positioning control,
measures have to be taken for the compensation of the inherent piezoelectric
effects such as creep and hysteresis and the compensation of varying external
mechanical loads and vibrations. By means of a non-contact inductive proximity
sensor attached to the bending actuator in combination with a high-accuracy
electronic circuit, the disturbing effects can be detected. The implementation
of such a sensor-actuator system into a closed-loop control allows a
compensation of negative effects, thus, high-efficient piezoceramics can be
used for high-accuracy positioning. In this paper, the focus is laid on the
electrodynamics of a specifically developed non-contact inductive proximity
sensor which is based on eddy current effects offering the possibility of
sensor integration in smallest spaces in combination with a high-accuracy
electronic circuit. By means of the theoretical formulation, a deep physical
understanding of the mechanism of the non-contact inductive proximity sensor is
given and provides the determination of the relative change of the inductance
as a function of the beam bender's tip deflection in an analytical way. The
electrodynamic approach is verified by experimental measurements. Since
piezoceramic bending actuators play an increasingly important role for energy
harvesting purposes, it is also shown, how tip deflection measurements in
combination with the presented non-contact inductive proximity sensor enable
the determination of the power dissipated in the transducer and thus, the
investigation and evaluation of the performance of piezoelectric energy
harvesting generators.}


\begin{document}
	
\section{Introduction}
\label{sec:1}

The piezoelectric effect combines electrical and mechanical parameters. The
conversion of a mechanical deformation into an electrical signal is called
\textit{direct piezoelectric effect}, the resulting deformation by applying an
electric field is referred to as the \textit{reciprocal piezoelectric effect}.
These effects can be used for piezoceramic sensors and actuators offering a
wide range of applications in the fields of acoustics, automation,
communication, health care, automotive and consumer products.

The development of low drift piezoceramics reduces the effects of relaxation
and hysteresis, but their low piezoelectric coefficients do not allow their use
as high-efficient
piezoceramic actuators. So, the range of applications with low drift
piezoceramics is limited due to small deflections and forces. For larger
deflections and forces a smart
sensor-actuator-system is required. This paper introduces a high efficient
piezoceramic bending actuator and an attached non-contact inductive proximity
sensor which can be implemented into a closed control loop. The usage of the
attached sensor in combination with a high-accuracy electronic circuit allows
the detection and compensation of disturbing influences like inherent
piezoelectric effects, external forces, moments and
vibrations~\cite{Multilayer}.

\section{Inductive Sensor Principle}
\label{sec:02}

\begin{figure}[h]
\centering
\includegraphics[scale=1]{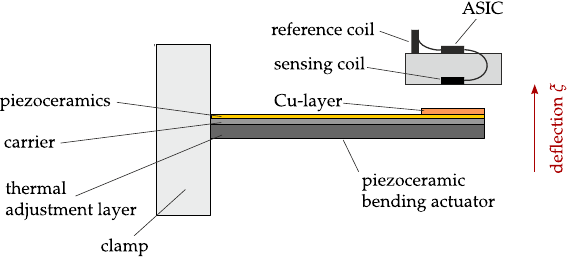}
\caption{Miniaturized non-contact inductive proximity
sensor to detect the tip deflection $\xi$ of a piezoelectric bending
actuator.}
\label{fig:Inductive01}
\end{figure}

Figure~\ref{fig:Inductive01} shows the measurement setup of the non-contact
inductive proximity sensor for tip deflection measurements of piezoelectric
bending actuators.

\begin{figure}[h]
\centering
\includegraphics[scale=1]{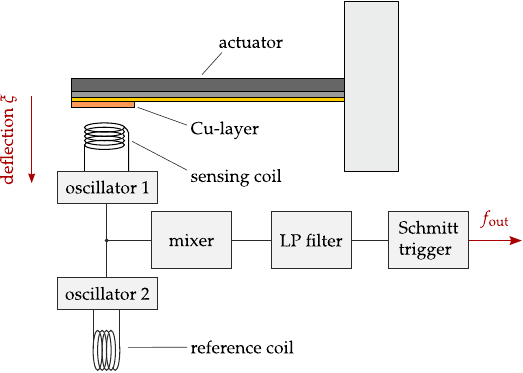}
\caption{Schematics of the inductive proximity sensor
electronics to detect the tip deflection $\xi$ of a piezoelectric bending
actuator.}
\label{fig:Inductive02}
\end{figure}

\noindent
The non-contact inductive proximity sensor consists of an application specific
integrated circuit (ASIC) and two external 3D-microcoils, which are connected
to the ASIC. One coil acts as sensing coil, the other one provides a reference.
The sensing coil is connected to the integrated oscillator $1$. Its resonant
frequency $f_{1}$ changes during the approach of the actuator's tip which is
covered by a detectable \ce{C}u-layer target (see Fig.~\ref{fig:Inductive02}).

If the sensing coil is driven by an alternating current $i_{1}$, an alternating
magnetic field will be generated. In its turn, this causes eddy currents within
the \ce{Cu}-layer generating magnetic fields, that counteract the external
magnetic
field (\textsc{Lenz}'s law). As a consequence, the external magnetic field
decreases, resulting in a decrease $\Delta L_{1}$ of the inductance $L_{1}$ of
the sensing coil. Thus, the changing resonant frequency of the oscillator $1$
is a measure for the deflection $\xi$ of the bender's tip. The reference coil
is connected to the integrated oscillator $2$. Its arrangement prevents an
influence by the \ce{Cu}-layer. With its constant inductance $L_{2}$, the
reference
coil provides the generation of the reference frequency $f_{2}$. The difference
$|f_{1}-f_{2}|$ and the sum $|f_{1}+f_{2}|$ of both frequency signals are
evaluated by means of the mixer and is fed through the following
lowpass filter. The lowpass filter suppresses deterministically caused
higher-frequency signals. The following Schmitt trigger provides a shaping of
both difference frequency
signals, thus the output signal $f_{\text{out}}$ is available in binary form
(TTL level)~\cite{Inductive}.

Both oscillator circuits consist of an integrated capacitance $C$ as well as an
active cell, maintaining the oscillation. The sensing and reference coil
complete the respective $L$-$C$-resonant circuit. In both $L$-$C$-resonant
circuits, the capacitance $C$ can be assumed to be equal. The frequency $f_{2}$
of the resonant circuit $2$ is specified by the reference
coil $L_{2}$ according to

\begin{equation}
f_{2}=\dfrac{1}{2 \uppi \sqrt{L_{2}C}}\text{.}
\label{Resonanzfrequenz Oszillator 2}
\end{equation}

\noindent
If the conductive Cu-layer target approaches the sensing coil, eddy currents
are generated within the layer weakening the alternating magnetic field of the
sensing coil. Thus, the effective inductance 
$L_{\text{eff}}=L_{1}-\Delta L_{1}$
of the sensing coil decreases and the oscillator's resonant frequency increases
from $f_{1}$ to

\begin{equation}
f_{\text{eff}}=\dfrac{1}{2 \uppi \sqrt{L_{\text{eff}}C}}=f_{1}
\sqrt{\dfrac{L_{1}}{L_{1}-\Delta L_{1}}}\text{.}
\label{Resonanzfrequenz Oszillator 1}
\end{equation}

\noindent
The general correlation between the relative frequency change and the relative
inductance change concerning the oscillator $1$ yields in combination with
$f_{\text{eff}}=f_{1}+ \Delta f_{1}$

\begin{equation}
\left| \dfrac{\Delta f_{1}}{f_{1}} \right|=\left| \dfrac{\left| \dfrac{\partial
f_{1}}{\partial L_{1}}\right| \Delta L_{1}}{f_{1}} \right|=\dfrac{1}{2}\left|
\dfrac{\Delta L_{1}}{L_{1}} \right| \text{.}
\label{relative Frequenzaenderung}
\end{equation}

\noindent
Since the relative inductance change $\Delta L_{1}$ amounts up to
$\SI{-30}{\percent}$ during approximation of the \ce{Cu}-layer target from a
distance of several centimeters to almost contact with the coil, the maximum
relative frequency change amounts $\SI{15}{\percent}$ with respect
to Eq.~(\ref{relative Frequenzaenderung}).
The mixer evaluates the difference of both frequencies $f_{\text{eff}}$ and
$f_{2}$. In order to achieve a reliable and a non-disturbed signal on TTL
level, the difference signal is filtered by a lowpass filter and shaped by a
Schmitt trigger. Thus, the output frequency results in

\begin{equation}
f_{\text{out}}=\left|f_{\text{eff}}-f_{2}\right| \text{.}
\label{Ausgangsfrequenz}
\end{equation}

\noindent
Dependent on $\Delta L_{1}/L_{1}$ of the oscillator $1$, the relative change of
the output frequency yields in combination with Eq.~(\ref{Ausgangsfrequenz}):

\begin{equation}
\left\vert 
\dfrac{\Delta f_{\text{out}}}{f_{\text{out}}}\right\vert =\left\vert
\dfrac{\left\vert \dfrac{%
\partial f_{\text{out}}}{\partial L_{1}}\right\vert _{\Delta L_{2}=0}\Delta
L_{1}}{%
f_{\text{out}}}\right\vert =\underset{\longrightarrow
15\,\%}{\underbrace{\dfrac{1}{2}%
\left\vert \dfrac{\Delta L_{1}}{L_{1}}\right\vert }}\underset{\gg 1}{%
\underbrace{\left\vert \dfrac{\sqrt{L_{2}}}{\sqrt{L_{1}}-\sqrt{L_{2}}}%
\right\vert }}
\label{Ausgangsfrequenz Aenderung}
\end{equation}

\noindent
Equation~(\ref{Ausgangsfrequenz Aenderung}) shows a very high relative output
frequency change being achieved, if both nominal inductances of the sensing and
reference coil are nearly
equal, even though the change of $L_{1}$ and thus, the distance change between
sensing coil and \ce{Cu}-layer target can be very small. This aspect
constitutes
the
choice of an inductive proximity sensor for high-accuracy tip deflection
measurements for a piezoelectric bending actuator.

\section{Equivalent Network Representation}
\label{sec:03}

\begin{figure}[h]
\centering
\includegraphics[scale=1]{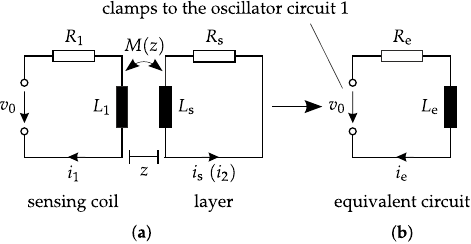}
\caption{Inductive interaction between sensing coil and
conductive layer: (\textbf{a}) Interconnection by means of the
distance-dependent mutual inductance $M(z)$ (transformer). (\textbf{b})
Equivalent circuit representation.}
\label{fig:Inductive03}
\end{figure}

The equivalent network of the sensing coil close to the conductive
\ce{Cu}-layer
is
illustrated in Fig.~\ref{fig:Inductive03}a~\cite{Proximity}. It is assumed,
that the Cu-layer carrying the eddy current $i_{\text{s}}$, can be represented
as a series connection of the inductance $L_{\text{s}}$ and the resistance
$R_{\text{s}}$. The sensing coil is represented by the inductance $L_{1}$ and
the resistance $R_{1}$. Due to the coupling between both $L$-$R$-circuits, an
alternate circuit representation is to be achieved, which allows for the
representation of the arising effective inductance $L_{\text{eff}}$ as a
function
of the distance $z$ between sensing coil and \ce{Cu}-layer (see
Fig.~\ref{fig:Inductive03}b).

\begin{figure}[h]
\centering
\includegraphics[scale=1]{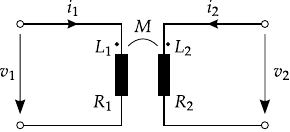}
\caption{General circuit representation of a transformer
without iron core and with coils wound in the same direction}
\label{fig:Inductive04}
\end{figure}

In order to describe the interaction between the \ce{Cu}-layer and sensing
coil,
an
equivalent circuit representation of a transformer without iron core is
considered (see Fig.~\ref{fig:Inductive04}). Assuming two coils with $N_{1}$
and $N_{2}$ turns and taking \textsc{Lenz}'s law into account, following pair
of equations can be formulated for both meshes of the circuit shown in 
Fig.~\ref{fig:Inductive04}:

\setlength{\abovedisplayskip}{-2pt}
\setlength\jot{0.3cm}
\begin{align}
-v_{1}+i_{1}R_{1}&=-N_{1}\dfrac{\mathrm{d}\varPhi_{1}}{\mathrm{d}t}
\label{voltage01}\\
-v_{2}+i_{2}R_{2}&=-N_{2}\dfrac{\mathrm{d}\varPhi_{2}}{\mathrm{d}t}
\label{voltage02}
\end{align}

\noindent
The quantities $\varPhi_{1}$ and $\varPhi_{2}$ correspond to the magnetic flux
correlated with coil $1$ and $2$, respectively. The individual fluxes are
proportional to the currents $i_{1}$ and $i_{2}$:

\setlength{\abovedisplayskip}{-2pt}
\setlength\jot{0.3cm}
\begin{align}
N_{1}\varPhi_{1}=L_{1}i_{1}+M\!\;i_{2}
\label{flux01}\\
N_{2}\varPhi_{2}=L_{2}i_{2}+M\!\;i_{1}
\label{flux02}
\end{align}

\noindent
In the Eqs.~(\ref{flux01}) and (\ref{flux02}) the quantities $L_{1}$ and
$L_{2}$
denote
the self-inductances, the quantity $M$ denotes the mutual inductance. Taking
Eqs.~(\ref{flux01}) and (\ref{flux02}) into consideration, both
Eqs.~(\ref{voltage01}) and (\ref{voltage02}) can be reformulated:

\setlength{\abovedisplayskip}{-2pt}
\setlength\jot{0.3cm}
\begin{align}
-v_{1}+i_{1}R_{1}=-L_{1}\dfrac{\mathrm{d}i_{1}}{\mathrm{d}t}-M\frac{\mathrm{d}i
_{2}}{\mathrm{d}t}
\label{flux03} \\
-v_{2}+i_{2}R_{2}=-L_{2}\dfrac{\mathrm{d}i_{2}}{\mathrm{d}t}-M\frac{\mathrm{d}i
_{1}}{\mathrm{d}t}
\label{flux04}
\end{align}

\noindent
The correlations formulated in Eqs.~(\ref{flux03}) and (\ref{flux04}) are
applied
to
the equivalent circuit representation in Fig.~\ref{fig:Inductive03}a. Following
analogies are achieved:

\begin{equation}
v_{1}=v_{0}\text{,}\hspace{5mm} i_{1}=i_{1}\text{,}\hspace{5mm}
i_{2}=i_{\text{s}}\text{,}\hspace{5mm} v_{2}=0
\label{analogies}
\end{equation}

\noindent
Using the analogies in Eq.~(\ref{analogies}), the Eqs.~(\ref{flux03}) and
(\ref{flux04}) can be reformulated according to

\setlength{\abovedisplayskip}{-2pt}
\setlength\jot{0.3cm}
\begin{align}
v_{0}=i_{1}R_{1}+L_{1}\dfrac{\mathrm{}i_{1}}{\mathrm{d}t}+M\frac{\mathrm{d}i_
{\text{s}}}{\mathrm{d}t}
\label{flux05} \\
0=i_{\text{s}}R_{\text{s}}+L_{\text{s}}\dfrac{\mathrm{d}i_{\text{s}}}{\mathrm{d
}t}+M\frac{\mathrm{d}i_{1}}{\mathrm{d}t}\text{.}
\label{flux06}
\end{align}

\noindent
In case of steady state currents and voltages, the complex root mean square
values yield:

\setlength{\abovedisplayskip}{-2pt}
\setlength\jot{0.3cm}
\begin{align}
\undersl{V}_{0}&=\left(R_{1}+\mathrm{j}\omega L_{1}\right)
\undersl{I}_{1}+\mathrm{j}\omega M\undersl{I}_{\text{s}}
\label{flux07} \\
0&=\left(R_{\text{s}}+\mathrm{j}\omega
L_{\text{s}}\right)
\undersl{I}_{\text{s}}+\mathrm{j}\omega M\undersl{I}_{1}
\label{flux08}
\end{align}

\noindent
Rearranging of Eq.~(\ref{flux08}) with respect to the eddy current
$\undersl{I}_{\text{s}}$ and following insertion into Eq.~(\ref{flux07})
results
in
the potential difference at the clamps of oscillator 1:

\begin{equation}
\undersl{V}_{0}=\undersl{I}_{1}\left[R_{1}+\dfrac{\omega^2M^2R_{\text{s}}}{R^2_
{\text{s}}+\omega^2L^2_{\text{s}}}+\mathrm{j}\omega\left(L_{1}-L_{\text{s}
}\dfrac{\omega^2M^2}{R^2_{\text{s}}+\omega^2L^2_{\text{s}}}\right)\right]
\label{flux09}
\end{equation}

\noindent
Due to the high conductivity of the Cu-layer target attached to the bender's
tip $(R_{\text{s}}^2 \ll \omega^2L^2_{\text{s}})$, Eq.~(\ref{flux09}) can be
simplified:

\begin{equation}
\undersl{V}_{0}=\undersl{I}_{1}\left[R_{1}+\dfrac{M^2}{L^2_{\text{s}}}R_{\text
{s}}+\mathrm{j}\omega\left(L_{1}-\dfrac{M^2}{L_{\text{s}}}\right)\right]
=\undersl{I}_{1}\left(R_{\text{e}}+\mathrm{j}\omega L_{\text{e}}\right)
\label{flux10}
\end{equation}

\noindent
Thus, the resistance $R_{\text{e}}$ and the inductance $L_{\text{e}}$ of the
equivalent network in Fig.~\ref{fig:Inductive03}b are known.
The definition of the coupling coefficient

\begin{equation}
k(z)=\dfrac{M(z)}{\sqrt{L_{1}L_{\text{s}}}}
\label{Kopplungsfaktor}
\end{equation}

\noindent
implies the formulation of the mutual inductance $M(z)$ as a function of
distance $z$ between the bender's tip and sensing coil. Thus, the equivalent
quantities $R_{\text{e}}$ and $L_{\text{e}}$ are dependent on the distance $z$
according to

\setlength{\abovedisplayskip}{-2pt}
\setlength\jot{0.3cm}
\begin{align}
R_{\text{e}}(z)&=R_{1}+R_{\text{s}}\dfrac{L_{1}}{L_{\text{s}}}k^2(z)
\shortintertext{and}
L_{\text{e}}(z)&=L_{1}\left[1-k^2(z)\right]\text{.}
\label{Le}
\end{align}

\noindent
Equation~(\ref{Le}) allows the formulation of the relative inductance change of
the sensing coil $1$:

\begin{equation}
\dfrac{L_{1}-L_{\text{e}}(z)}{L_{1}}=\frac{\Delta L_{1}}{L_{1}}=k^2(z)
\label{Induktivitaetsaenderung}
\end{equation}

\noindent
With respect to Eq.~(\ref{Resonanzfrequenz Oszillator 2}), the basic inductance
$L_{1}(z \rightarrow \infty)$ is correlated with the frequency $f_{1}$ of
oscillator $1$:

\begin{equation}
L_{1}=\dfrac{1}{4 \uppi^2 f^2_{1}C}
\label{L1}
\end{equation}

\noindent
Similarly, the equivalent inductance yields

\begin{equation}
L_{\text{e}}(z)=\dfrac{1}{4 \uppi^2 f^2_{\text{e}}(z)C}\text{.}
\label{Leq}
\end{equation}

\noindent
Thus, Eq.~(\ref{Induktivitaetsaenderung}) can be reformulated:

\begin{equation}
\frac{\Delta L_{1}}{L_{1}}=1-\left(\frac{f_{1}}{f_{\text{e}}(z)}\right)^2
\label{rel}
\end{equation}

\noindent
In a next step, the distance-dependent relative inductance change will be
calculated by an electrodynamic approach.

\section{Inductance of a Circular Loop Influenced by a Conductive Layer}
\label{sec:04}

\begin{figure}[h]
\centering
\includegraphics[scale=1]{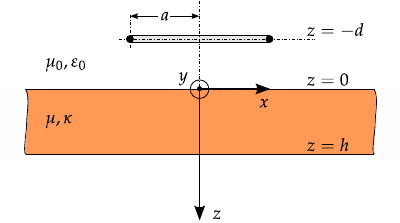}
\caption[Inductive sensor 05]{Arrangement of a circular loop opposite to a
conductive layer (side view)}
\label{fig:Inductive05}
\end{figure}

The arrangement of a circular loop opposite to a conductive layer is shown in
Fig.~\ref{fig:Inductive05}. With respect to the origin of the coordinate
system, the circular loop is parallel to the surface layer in distance $z=-d$.
The represented arrangement is axially symmetric, thus it is described by means
of cylindrical coordinates $(r,\theta,z)$.

With respect to \textsc{Maxwell}'s equations for time-variable fields, the
electric field $\mathbf{E}$ generates a magnetic field $\mathbf{H}$ by means of
the conduction and displacement current. On its part, the magnetic field
$\mathbf{H}$ generates an electric field by means of induction. Thus, electric
and magnetic fields are connected with each other. For the physical description
of the arrangement illustrated in Fig.~\ref{fig:Inductive05}, the vector of
the magnetic field strength $\mathbf{H}(\mathbf{r},t)$ is of interest.

\subsection{Solution Approach}
\label{ssec:01}

The fundamental describing \textsc{Maxwell} equations are:

\setlength{\abovedisplayskip}{-2pt}
\setlength\jot{0.3cm}
\begin{align}
\text{curl}\;\!\mathbf{E}(\mathbf{r},t)&=-\dfrac{\partial\mathbf{B}(\mathbf{r}
,t)}{\partial t}
\label{rotE} \\
\text{curl}\;\!\mathbf{H}(\mathbf{r},t)&=\dfrac{\partial\mathbf{D}(\mathbf{r},t
)}{\partial t}+\mathbf{J}(\mathbf{r},t)
\label{rotH}
\end{align}

\noindent
Taking the material-specific relations for the electric displacement
$\mathbf{D}(\mathbf{r},t)$, the current density $\mathbf{J}(\mathbf{r},t)$ and
the magnetic induction $\mathbf{B}(\mathbf{r},t)$ in accordance with

\setlength{\abovedisplayskip}{-2pt}
\setlength\jot{0.3cm}
\begin{align}
\mathbf{D}(\mathbf{r},t)&=\varepsilon \mathbf{E}(\mathbf{r},t)
\label{D} \\
\mathbf{J}(\mathbf{r},t)&=\kappa \mathbf{E}(\mathbf{r},t)
\label{J} \\
\mathbf{B}(\mathbf{r},t)&=\mu \mathbf{H}(\mathbf{r},t)
\label{B}
\end{align}

\noindent
into account, it can be written:

\setlength{\abovedisplayskip}{-2pt}
\setlength\jot{0.3cm}
\begin{align}
\text{curl}\;\!\mathbf{E}(\mathbf{r},t)&=-\mu
\dfrac{\partial\mathbf{H}(\mathbf{r},t)}{\partial t}
\label{rotE1} \\
\text{curl}\;\!\mathbf{H}(\mathbf{r},t)&=\varepsilon
\dfrac{\partial\mathbf{E}(\mathbf{r},t)}{\partial t}+\kappa
\mathbf{E}(\mathbf{r},t)
\label{rotH1}
\end{align}

\noindent
The permittivity $\varepsilon$, the conductivity $\kappa$ and the permeability
$\mu$ in Eqs.~(\ref{rotE1}) and (\ref{rotH1}) are assumed to exhibit a
time-independent, isotropic behavior. The application of the vector operation
"$\mathrm{curl}$" to Eq.~(\ref{rotH1}) and following insertion of
Eq.~(\ref{rotE1}) yields:

\begin{equation}
\text{curl}\;\!\text{curl}\;\!\mathbf{H}(\mathbf{r},t)=-\varepsilon \mu
\dfrac{\partial^2\mathbf{H}(\mathbf{r},t)}{\partial t^2}-\kappa \mu
\dfrac{\partial \mathbf{H}(\mathbf{r},t)}{\partial t}
\label{rotrot}
\end{equation}

\noindent
According to the calculation rules of vector analysis, it can be written in
general~\cite{Mathematics}:

\begin{equation}
\text{curl}\;\!\text{curl}\;\!\mathbf{X}=\text{grad}\;\!\text{div}\;\!\mathbf{X
}-\Delta\mathbf{X}
\label{rotrot2}
\end{equation}

\noindent
Thus, Eq.~(\ref{rotrot}) can be reformulated:

\begin{equation}
\text{grad}\;\!\text{div}\;\!\mathbf{H}(\mathbf{r},t)-\Delta\mathbf{H}(\mathbf
{r},t)=-\varepsilon \mu \dfrac{\partial^2\mathbf{H}(\mathbf{r},t)}{\partial
t^2}-\kappa \mu \dfrac{\partial
\mathbf{H}(\mathbf{r},t)}{\partial t}
\label{rotrot3}
\end{equation}

\noindent
Since there is no source of magnetic field
$(\text{div}\;\!\mathbf{H}(\mathbf{r},t)=0)$, Eq.~(\ref{rotrot3}) can be
simplified:

\begin{equation}
\Delta\mathbf{H}(\mathbf{r},t)-\varepsilon \mu
\dfrac{\partial^2\mathbf{H}(\mathbf{r},t)}{\partial t^2}-\kappa \mu
\dfrac{\partial \mathbf{H}(\mathbf{r},t)}{\partial t}=\mathbf{0}
\label{Laplace}
\end{equation}

\noindent
Equation~(\ref{Laplace}) results in two differential equations, one for the
\textit{metallic semi-infinite space} $(\varepsilon=0,\,\mu,\,\kappa)$
and one
for the \textit{medium air} $(\varepsilon=\varepsilon_{0},\, \mu=\mu_{0},\,
\kappa=0)$:

\setlength{\abovedisplayskip}{-2pt}
\setlength\jot{0.3cm}
\begin{align}
\Delta\mathbf{H}(\mathbf{r},t)-\kappa \mu \dfrac{\partial
\mathbf{H}(\mathbf{r},t)}{\partial t}&=\mathbf{0}\hspace{15mm}\text{metal}
\label{Metall} \\
\Delta\mathbf{H}(\mathbf{r},t)-\varepsilon_{0} \mu_{0}
\dfrac{\partial^2\mathbf{H}(\mathbf{r},t)}{\partial
t^2}&=\mathbf{0}\hspace{15mm}\text{air}
\label{Luft}
\end{align}

\noindent
The expressions~(\ref{Metall}) and (\ref{Luft}) represent the electromagnetic
skin and wave equation concerning the magnetic field in metal and air,
respectively. Since cylindrical coordinates are used, the \textit{Laplacian}
$\Delta$ has to be formulated explicitly. Taking the angle independence of the
magnetic field $(\partial \mathbf{H}(\mathbf{r},t)/\partial \theta=0)$ into
consideration, the necessary differential equations of the $z$-component of the
magnetic field result in their final formulation~\cite{Impedance}:

\setlength{\abovedisplayskip}{-2pt}
\setlength\jot{0.3cm}
\begin{align}
\dfrac{1}{r}\dfrac{\partial}{\partial r}(r \dfrac{\partial H_{z}}{\partial
r})+\dfrac{\partial^2 H_{z}}{\partial z^2}-\kappa \mu \dfrac{\partial
H_{z}}{\partial t}&=0
\hspace{10mm}\text{metal}
\label{Metall2} \\
\dfrac{1}{r}\dfrac{\partial}{\partial r}(r \dfrac{\partial H_{z}}{\partial
r})+\dfrac{\partial^2 H_{z}}{\partial z^2}-\varepsilon_{0} \mu_{0}
\dfrac{\partial^2 H_{z}}{\partial t^2}&=0
\hspace{10mm}\text{air}
\label{Luft2}
\end{align}

\subsection{Alternating Magnetic Field Within a Conductive Layer}
\label{ssec:02}

In order to find a solution for the alternating magnetic field within the
conductive layer, Eq.~(\ref{Metall2}) is reformulated according to:

\begin{equation}
\dfrac{\partial^2 H_{z}}{\partial r^2}+\dfrac{\partial^2 H_{z}}{\partial
z^2}+\dfrac{1}{r}\dfrac{\partial H_{z}}{\partial r}-\kappa \mu \dfrac{\partial
H_{z}}{\partial t}=0
\label{Metall3}
\end{equation}

\noindent
The separation of variables serves as method for finding a general solution:

\begin{equation}
\undersl{H}_{z}(r,z,t)=V(r,z)W(t)
\label{Metall4}
\end{equation}

\noindent
Since the circular loop is subjected to a sinusoidal alternating current
$\undersl{I}(t)=\undersl{I}\mathrm{e}^{\;\!\mathrm{j}\omega t}$, the real time
function $H_{z}(r,z,t)$ is replaced by the complex time function
$\undersl{H}_{z}(r,z,t)$. Thus, Eq.~(\ref{Metall4}) can be written as follows:

\begin{equation}
\undersl{H}_{z}(r,z,t)=\undersl{V}(r,z)\mathrm{e}^{\;\!\mathrm{j}\omega t}
\label{Metall5}
\end{equation}

\noindent
Insertion of Eq.~(\ref{Metall5}) into Eq.~(\ref{Metall3}) and following
separation
of
the variables $(r,z,t)$ yields

\begin{equation}
\dfrac{1}{\undersl{V}}\left(\dfrac{\partial^2 \undersl{V}}{\partial
r^2}+\dfrac{\partial^2 \undersl{V}}{\partial z^2}+\dfrac{1}{r}\dfrac{\partial
\undersl{V}}{\partial r}\right)=\mathrm{j} \omega \kappa \mu=:-\nu^2\text{.}
\label{Metall6}
\end{equation}

\noindent
Here, $\nu$ is a constant, since the left side of Eq.~(\ref{Metall6}) is
independent of $t$ and the right side is independent of $r$ and $z$. Applying
the method of separation of variables again results in

\begin{equation}
\undersl{V}(r,z)=\undersl{X}(r)\undersl{Y}(z)\text{.}
\label{Metall7}
\end{equation}

\noindent
Insertion of Eq.~(\ref{Metall7}) into Eq.~(\ref{Metall6}) yields

\begin{equation}
\dfrac{1}{\undersl{X}(r)}\dfrac{\mathrm{d}^2
\undersl{X}(r)}{\mathrm{d}r^2}+\dfrac{1}{\undersl{X}(r)}\dfrac{1}{r}\dfrac
{\mathrm{d}
\undersl{X}(r)}{\mathrm{d}r}=-\dfrac{1}{\undersl{Y}(z)}\dfrac{\mathrm{d}^2
\undersl{Y}(z)}{\mathrm{d}z^2}-\nu^2=:-u_{n}^2\text{,}
\label{Metall8}
\end{equation}

\noindent
where $u_{n}$ denotes another constant. Two ordinary differential equations for
$\undersl{X}(r)$ and $\undersl{Y}(z)$ result from Eq.~(\ref{Metall8}):

\setlength{\abovedisplayskip}{-2pt}
\setlength\jot{0.3cm}
\begin{align}
\dfrac{\mathrm{d}^2\undersl{Y}(z)}{\mathrm{d}z^2}-\undersl{Y}(z)(u_{n}^2-\nu^2)
=0
\label{Metall9} \\
\dfrac{\mathrm{d}^2\undersl{X}(r)}{\mathrm{d}r^2}+\dfrac{1}{r}\dfrac{\undersl{X
}(r)}{\mathrm{d}r}+\undersl{X}(r)u_{n}^2=0
\label{Metall10}
\end{align}

\noindent
The solution $\undersl{Y}(z)$ of the homogeneous linear differential
equation~(\ref{Metall9}) results from the linear combination of two linear
independent
solutions (basis solutions). Here, the solutions $\mathrm{e}^{\lambda z}$ and
$\mathrm{e}^{-\lambda z}$ represent the fundamental system of the differential
equation. Following solution is achieved:

\begin{equation}
\undersl{Y}(z)=\undersl{A}_{1n}\mathrm{e}^{-z\sqrt{u_{n}^2+\mathrm{j} \omega
\kappa \mu}}+\undersl{A}_{2n}\mathrm{e}^{z\sqrt{u_{n}^2+\mathrm{j} \omega
\kappa \mu}}
\label{Metall11}
\end{equation}

\noindent
In order to find a solution $\undersl{X}(r)$ for the differential
equation~(\ref{Metall10}), a transition to the variable $\rho:=u_{n}r$ is
defined. With
the definition

\begin{equation}
\undersl{Z}(\rho):=\undersl{X}(\rho/u_{n})
\label{Metall12}
\end{equation}

\noindent
and following insertion into Eq.~(\ref{Metall10}), \textsc{Bessel}'s
differential
equation follows:

\begin{equation}
\dfrac{\mathrm{d}^2\undersl{Z}(\rho)}{\mathrm{d}\rho^2}+\dfrac{1}{\rho}\dfrac
{\mathrm{d}\undersl{Z}(\rho)}{\mathrm{d}\rho}+\undersl{Z}(\rho)=0
\label{Bessel}
\end{equation}

\newpage
\noindent
The general solution of Eq.~(\ref{Bessel}) results in~\cite{Vektoranalysis}:

\begin{equation}
\undersl{Z}(\rho)=\undersl{C}_{1}J_{0}(\rho)
\label{Bessel1}
\end{equation}

\noindent
Choosing $\undersl{C}_{1}=1$, which does not influence the total solution at
all, and taking Eq.~(\ref{Metall12}) into consideration, the solution
$\undersl{X}(r)$ is achieved:

\begin{equation}
\undersl{X}(r)=J_{0}(u_{n}r)\text{.}
\label{Bessel2}
\end{equation}

\noindent
The expression $J_{0}(u_{n}r)$ represents \textsc{Bessel}'s function of
\textit{first kind} and \textit{order zero}. Taking Eqs.~(\ref{Metall5}) and
(\ref{Metall3}) into account, the formal solution is achieved by
superposition~\cite{Vektoranalysis}:

\begin{equation}
\undersl{H}_{z}(r,z,t)=\,\,\mathrm{e}^{\;\!\mathrm{j}\omega t}
\sum\limits_{n=1}^{\infty}J_{0}(u_{n}r) \undersl{A}_{1n} \mathrm{e}^{-z
\sqrt{u_{n}^2+\mathrm{j} \omega \kappa \mu}}+\mathrm{e}^{\;\!\mathrm{j}\omega
t} \sum\limits_{n=1}^{\infty}J_{0}(u_{n}r) \undersl{A}_{2n} \mathrm{e}^{z
\sqrt{u_{n}^2+\mathrm{j} \omega \kappa \mu}}
\label{Bessel3}
\end{equation}

\noindent
Due to the continuous spectrum of the eigenvalues $u$, Eq.~(\ref{Bessel3}) can
be
written in integral form~\cite{Electromagnetics}:

\begin{equation}
\undersl{H}_{z}(r,z,t)=\,\,\mathrm{e}^{\;\!\mathrm{j}\omega t}
\int\limits_{0}^{\infty}J_{0}(ur) \undersl{A}_{1} \mathrm{e}^{-z
\sqrt{u^2+\mathrm{j} \omega \kappa \mu}}\,
\mathrm{d}u+\mathrm{e}^{\;\!\mathrm{j}\omega t}
\int\limits_{0}^{\infty}J_{0}(ur) \undersl{A}_{2} \mathrm{e}^{z
\sqrt{u^2+\mathrm{j} \omega \kappa \mu}}\, \mathrm{d}u
\label{Loesung Metall H1}
\end{equation}

\noindent
Finally, the quantities $\undersl{A}_{1}$ and $\undersl{A}_{2}$ can be solved
by means of corresponding boundary conditions.
\\

The general solution for the alternating magnetic field in the air is similar
to Eq.~(\ref{Loesung Metall H1}), while the term $\mathrm{j}\omega \kappa\mu$
has
to be replaced by the term $-\varepsilon_{0} \mu_{0} \omega^2$. Since the
oscillator $1$ of the available inductive proximity sensor exhibits an
oscillation frequency of $f\approx\SI{17}{\mega\hertz}$, the term
$-\varepsilon_{0} \mu_{0} \omega^2$ can be neglected. Thus, the alternating
magnetic field in the air is formulated as following:

\begin{equation}
\undersl{H}_{z}(r,z,t)=\,\,\mathrm{e}^{\;\!\mathrm{j}\omega t}
\int\limits_{0}^{\infty}J_{0}(ur)
\left[\undersl{B}_{1}\mathrm{e}^{-zu}+\undersl{B}_{2}
\mathrm{e}^{\;\!zu}\right] \, \mathrm{d}u
\label{Bessel4}
\end{equation}

\subsection{Magnetic Field of a Circular Loop}
\label{ssec:03}

\begin{figure}[h]
\centering
\includegraphics[scale=1]{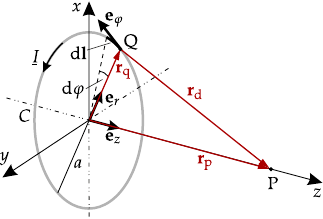}
\caption{Geometry of a circular loop for the calculation of
the magnetic field along the $z$-axis.}
\label{fig:Inductive6}
\end{figure}

The geometrical quantities being necessary for the calculation of the magnetic
field of a circular loop (radius $a$) along its axis in distance $z$ from the
center (point of origin) are shown in Fig.~\ref{fig:Inductive6}. The vector
$\mathrm{d}\mathbf{l}$ points in direction of the alternating current
$\undersl{I}(t)=\undersl{I}\mathrm{e}^{\;\!\mathrm{j}\omega t}$. The current
loop in space is characterized by the quantity $C$, the vector
$\mathbf{r}_{\mathrm{d}}$ points from the source point $\mathrm{Q}$ to the
space point $\mathrm{P}$. Due to the present symmetry, cylindrical coordinates
are used for following considerations.

\noindent
In order to calculate the magnetic field of a circular loop along its axis of
symmetry, the \textsc{Biot}-\textsc{Savart} law is considered:

\begin{equation}
\underslv{\mathbf{H}}(\mathbf{r}_{\mathrm{p}},t)=\undersl{I}\mathrm{e}^{\;\
!\mathrm{j}\omega t}\int\limits_{C} \dfrac{\mathrm{d}\mathbf{l} \times
\mathbf{r}_{\mathrm{d}}}{r^3_{\mathrm{d}}}
\label{Biot}
\end{equation}

\noindent
The vector $\mathrm{d}\mathbf{l}$ can be defined using the unit vector
$\mathbf{e}_{\varphi}$:

\begin{equation}
\mathrm{d}\mathbf{l}=a\;\!\mathrm{d}\varphi\;\!\mathbf{e}_{\varphi}
\label{Biot1}
\end{equation}

\noindent
The vector $\mathbf{r}_{\mathrm{d}}$ pointing from source point $\mathrm{Q}$ at
space point $\mathrm{P}$ results from the difference of the position
vectors $\mathbf{r}_{\mathrm{p}}$ and $\mathbf{r}_{\mathrm{q}}$:

\begin{equation}
\mathbf{r}_{\mathrm{d}}=\mathbf{r}_{\mathrm{p}}-\mathbf{r}_{\mathrm{q}}
=z\mathbf{e}_{z}-a\mathbf{e}_{r}
\label{Biot2}
\end{equation}

\noindent
The vector product $\mathrm{d}\mathbf{l} \times \mathbf{r}_{\mathrm{d}}$  can
be determined with Eqs.~(\ref{Biot1}) and (\ref{Biot2}).

\begin{equation}
\mathrm{d}\mathbf{l} \times
\mathbf{r}_{\mathrm{d}}=a\,\mathrm{d}\varphi\mathbf{e}_{\varphi}\times
(z\mathbf{e}_{z}+a\mathbf{e}_{r})=za\,\mathrm{d}\varphi
\mathbf{e}_{r}+a^2\,\mathrm{d}\varphi \mathbf{e}_{z}
\label{Biot3}
\end{equation}

\noindent
Furthermore, Eq.~(\ref{Biot2}) yields

\begin{equation}
\left|\mathbf{r}_{\mathrm{d}}\right|=\sqrt{z^2+a^2}\text{.}
\label{Biot4}
\end{equation}

\noindent
Insertion of Eqs.~(\ref{Biot3}) and (\ref{Biot4}) into Eq.(\ref{Biot}) yields
the
axial
component of the magnetic field:

\begin{equation}
\undersl{H}_{z}(r=0,z,t)=\dfrac{a^2\undersl{I}}{2(z^2+a^2)^{3/2}}\mathrm{e}^{\
;\!\mathrm{j}\omega t}
\label{Biot5}
\end{equation}

\noindent
In order to provide a general representation of the magnetic field of a
circular loop in the following form

\begin{equation}
\undersl{H}_{z}(r,z,t)=\,\,\mathrm{e}^{\;\!\mathrm{j}\omega t}
\int\limits_{0}^{\infty}J_{0}(ur) \left[\undersl{B}_{1}
\mathrm{e}^{-zu}+\undersl{B}_{2} \mathrm{e}^{zu}\right] \, \mathrm{d}u\text{,}
\label{Biot6}
\end{equation}

\noindent
following integral has to be evaluated~\cite{Loss}:

\begin{equation}
\int\limits_{0}^{\infty}\mathrm{e}^{-zu}J_{0}(au) \, \mathrm{d}u
\label{Biot7}
\end{equation}

\noindent
\textsc{Bessel}'s function of \textit{first kind} and \textit{order zero} can
be defined by a complex exponential function (\textit{generating function}) in
integral form~\cite{Bessel,Physicists}:

\begin{equation}
J_{0}(au)=\dfrac{1}{2\uppi}\int\limits_{0}^{2\uppi}
\mathrm{e}^{\;\!\mathrm{j}au\cos \varphi}\, \mathrm{d}\varphi
\label{Biot8}
\end{equation}

\noindent
Insertion of Eq.~(\ref{Biot8}) into Eq.~(\ref{Biot7}) results in

\begin{equation}
\dfrac{1}{2\uppi}\int\limits_{0}^{2\uppi}\int\limits_{0}^{\infty}
\mathrm{e}^{-(z-\mathrm{j}a\cos \varphi)u}\, \mathrm{d}u =
\dfrac{1}{2\uppi}\int\limits_{0}^{2\uppi}
\underbrace{\dfrac{1}{z-\mathrm{j}a\cos \varphi}}_{\substack{(*)}} \,
\mathrm{d}\varphi \text{.}
\label{Biot9}
\end{equation}

\noindent
Rewriting $(*)$ in the form

\begin{equation}
\dfrac{1}{z-\mathrm{j}a\cos \varphi}=\dfrac{z}{z^2+a^2 \cos^2\varphi}
+\mathrm{j}\dfrac{a \cos \varphi}{z^2+a^2 \cos^2\varphi}\text{,}
\label{Biot10}
\end{equation}

\noindent
inserting it into Eq.~(\ref{Biot9}) and integrating it in combination with
appropriate integral tables~\cite{Taschenbuch} gives following correlation

\begin{equation}
\int\limits_{0}^{\infty}\mathrm{e}^{-zu}J_{0}(au) \,
\mathrm{d}u=\dfrac{1}{\sqrt{z^2+a^2}}\text{.}
\label{Biot11}
\end{equation}

\noindent
The derivative with respect to the variable $a$ according to

\begin{equation}
\int\limits_{0}^{\infty}\mathrm{e}^{-zu}\dfrac{\mathrm{d}}{\mathrm{d}a}J_{0}(au
) \, \mathrm{d}u=\dfrac{\mathrm{d}}{\mathrm{d}a}\dfrac{1}{\sqrt{z^2+a^2}}
\label{Biot12}
\end{equation}

\noindent
together with the general derivation rule for \textsc{Bessel}'s
function~\cite{Sciences}

\begin{equation}
\dfrac{\mathrm{d}}{\mathrm{d}x}J_{p}(x)=\dfrac{p}{x}J_{p}(x)-J_{p+1}(x)
\label{Biot13}
\end{equation}

\noindent
results in the expression

\begin{equation}
\int\limits_{0}^{\infty}\mathrm{e}^{-zu}J_{1}(au)u \,
\mathrm{d}u=\dfrac{a}{(z^2+a^2)^{3/2}}\text{.}
\label{Biot14}
\end{equation}

\noindent
Multiplication with $\undersl{I}\mathrm{e}^{\;\!\mathrm{j}\omega t}a/2$ on both
sides of Eq.~(\ref{Biot14}) finally results in

\begin{equation}
\dfrac{1}{2}\mathrm{e}^{\;\!\mathrm{j}\omega
t}\int\limits_{0}^{\infty}\mathrm{e}^{-zu}J_{1}(au)au\undersl{I} \,
\mathrm{d}u=\dfrac{a^2\undersl{I}}{2(z^2+a^2)^{3/2}}\mathrm{e}^{\;\!\mathrm{j
}\omega t}\text{,}
\label{Biot15}
\end{equation}

\noindent
which exactly matches the analytical description (\ref{Biot5}) of the magnetic
field of a circular loop along the $z$-axis.
\\

Since the origin of the coordinate system in Fig.~\ref{fig:Inductive05} is
not located in the center of the circular loop, but shifted in $z$-direction
$(z=-d)$, Eq.~(\ref{Biot15}) results in

\begin{equation}
\undersl{H}_{z}(r=0,z,t)=\dfrac{1}{2}\mathrm{e}^{\;\!\mathrm{j}\omega t}
\int\limits_{0}^{\infty}\mathrm{e}^{-(z+d)u}J_{1}(au) au \undersl{I} \,
\mathrm{d}u=\dfrac{1}{2}\mathrm{e}^{\;\!\mathrm{j}\omega t}
\int\limits_{0}^{\infty}\mathrm{e}^{-zu}\mathrm{e}^{-du}J_{1}(au) au
\undersl{I} \, \mathrm{d}u \text{.}
\label{Biot16}
\end{equation}

\noindent
The expression defined in Eq.~(\ref{Biot16}) has to be equivalent to the
general
formulation of the magnetic field in Eq.~(\ref{Biot6}) with respect to $r=0$,
thus
with $J_{0}(0)=1$ following condition can be formulated:

\begin{equation}
\int\limits_{0}^{\infty} \left[\undersl{B}_{1} \mathrm{e}^{-zu}+\undersl{B}_{2}
\mathrm{e}^{zu}\right] \, \mathrm{d}u=
\dfrac{1}{2}\int\limits_{0}^{\infty}\mathrm{e}^{-zu}\mathrm{e}^{-du}J_{1}(au)
au \undersl{I} \, \mathrm{d}u
\label{Biot17}
\end{equation}

\noindent
Comparing left and right side of Eq.~(\ref{Biot17}) immediately yields

\begin{equation}
\undersl{B}_{2}=0 \text{,}
\label{Biot18}
\end{equation}

\noindent
resulting in

\begin{equation}
\int\limits_{0}^{\infty} \undersl{B}_{1} \mathrm{e}^{-zu} \,
\mathrm{d}u=\dfrac{1}{2}\int\limits_{0}^{\infty}\mathrm{e}^{-zu}\mathrm{e}^{-du
}J_{1}(au) au \undersl{I} \, \mathrm{d}u \text{.}
\label{Biot19}
\end{equation}

\noindent
Comparig left and right side of Eq.~(\ref{Biot19}), the constant
$\undersl{B}_{1}$
can be formulated according to

\begin{equation}
\undersl{B}_{1}=\dfrac{1}{2} au\undersl{I}\mathrm{e}^{-du}J_{1}(au) \,
\mathrm{d}u \text{.}
\label{Biot20}
\end{equation}

\noindent
Inserting Eq.~(\ref{Biot20}) into Eq.~(\ref{Biot6}) and taking
Eq.~(\ref{Biot18})
into
consideration, the general description of the magnetic field of a circular loop
dependent on the coordinates $(r,z,t)$ is achieved:

\begin{equation}
\undersl{H}_{z}(r,z,t)=\mathrm{e}^{\;\!\mathrm{j}\omega
t}\int\limits_{0}^{\infty}\undersl{B}_{1}\mathrm{e}^{-zu}J_{0}(ur) \,
\mathrm{d}u \hspace{5mm} (z>-d)
\label{Biot21}
\end{equation}

\noindent
Equation~(\ref{Biot21}) allows for calculation of the magnetic field in the
plane of the circular loop ($z=-d$).

\begin{equation}
\undersl{H}_{\text{coil}}=\undersl{H}_{z}(r,z=-d,t)=\mathrm{e}^{\;\!\mathrm{j
}\omega t}\int\limits_{0}^{\infty}\undersl{B}_{1}\mathrm{e}^{ud}J_{0}(ur) \,
\mathrm{d}u
\label{Biot22}
\end{equation}

\noindent
The achieved expression in Eq.~(\ref{Biot22}) will play an important role for
the
analytical calculation of the relative inductance change of a circular loop in
presence of a conductive layer.

\subsection{Influence of a Conductive Layer}
\label{ssec:04}

In order to investigate the influence of a conductive layer of thickness $h$ on
a circular loop, the upper and lower surface of the conductive layer are
assumed to be formed by the two planes at $z=0$ and $z=h$ (see Fig.~
\ref{fig:Inductive05}). For the consideration of the influence on the space
$z<0$, the magnetic field of the circular loop Eq.~(\ref{Biot21}) has to be
modified by an additional term. This will ensure  the analytical formulation
corresponding to the general form in Eq.~(\ref{Biot6}) and moreover, the part
of
the additional magnetic field is zero for $z\rightarrow -\infty$. The magnetic
field $\undersl{H}_{z}^{\prime}$ between circular loop and conductive layer
$(-d<z<0)$ can be described by

\begin{equation}
\undersl{H}_{z}^{\prime}(r,z,t)=\mathrm{e}^{\;\!\mathrm{j}\omega t}
\int\limits_{0}^{\infty}J_{0}(ur) \left[\undersl{B}_{1}
\mathrm{e}^{-zu}+\undersl{B}_{2} \mathrm{e}^{zu}\right] \, \mathrm{d}u\text{,}
\label{Biot23}
\end{equation}

\noindent
where $\undersl{B}_{1}$ has already been defined by Eq.~(\ref{Biot20}).
$\undersl{B}_{2}$ has to be determined.
\\

Within the conductive layer, the description of the magnetic field results from
Eq.~(\ref{Loesung Metall H1}). Using the definition

\begin{equation}
k=\dfrac{\omega \kappa \mu}{u^2}\text{,}
\label{k}
\end{equation}

\noindent
for $(0<z<h)$, expression~(\ref{Loesung Metall H1}) is reformulated according
to

\begin{equation}
\undersl{H}_{z}^{\prime}(r,z,t)=\,\,\mathrm{e}^{\;\!\mathrm{j}\omega t}
\int\limits_{0}^{\infty}J_{0}(ur) \undersl{A}_{1}
\mathrm{e}^{-uz\sqrt{1+\mathrm{j} k}}\,
\mathrm{d}u+\mathrm{e}^{\;\!\mathrm{j}\omega t}
\int\limits_{0}^{\infty}J_{0}(ur) \undersl{A}_{2} \mathrm{e}^{uz
\sqrt{1+\mathrm{j} k}}\, \mathrm{d}u\text{.}
\label{Loesung Metall H}
\end{equation}

\noindent
The magnetic field outside the layer $(z>h)$ can be described similarly to the
expression~(\ref{Biot6}). However, the positive exponential term has to be
excluded:

\begin{equation}
\undersl{H}_{z}^{\prime}(r,z,t)=\mathrm{e}^{\;\!\mathrm{j}\omega
t}\int\limits_{0}^{\infty}\undersl{C}_{1}\mathrm{e}^{-zu}J_{0}(ur) \,
\mathrm{d}u \hspace{5mm} (z>h)
\label{Biot24}
\end{equation}

\noindent
In order to determine the four unknown quantities $\undersl{B}_{1}$,
$\undersl{A}_{1}$, $\undersl{A}_{2}$ and $\undersl{C}_{1}$, the continuity
characteristics of the normal component $B_{\mathrm{n}}$ of the magnetic
induction and the discontinuity characteristics of the normal component
$H_{\mathrm{n}}$ of the magnetic field are consulted with respect to both
boundary surfaces of the conductive layer at $z=0$, $z=h$ and $r=0$, i.e.

\begin{equation}
B_{\mathrm{n}1}=B_{\mathrm{n}2}
\label{normalB}
\end{equation}

\noindent
and
\begin{equation}
\dfrac{\mathrm{d}H_{\mathrm{n}1}}{\mathrm{d}z}=\dfrac{\mathrm{d}H_{\mathrm{n}2}
}{\mathrm{d}z}\text{.}
\label{normalH}
\end{equation}

\newpage
\noindent
The following system of equations is achieved:

\setlength{\abovedisplayskip}{-2pt}
\setlength\jot{0.3cm}
\begin{align}
\mu_0(\undersl{B}_{1}+\undersl{B}_{2})&=\mu(\undersl{A}_{1}+\undersl{A}_{2})
\label{Stetigkeit 1} \\
\undersl{B}_{1}-\undersl{B}_{2}&=\sqrt{1+\mathrm{j}k}(\undersl{A}_{1}-\undersl
{A}_{2})
\label{Stetigkeit 2} \\
\mu(\undersl{A}_{1}\mathrm{e}^{-uh\sqrt{1+\mathrm{j}k}}+\undersl{A}_{2}\mathrm
{e}^{uh\sqrt{1+\mathrm{j}k}})&=\mu_{0}\undersl{C}_{1}\mathrm{e}^{-uh}
\label{Stetigkeit 3} \\
\sqrt{1+\mathrm{j}k}(\undersl{A}_{1}\mathrm{e}^{-uh\sqrt{1+\mathrm{j}k}}
-\undersl{A}_{2}\mathrm{e}^{uh\sqrt{1+\mathrm{j}k}})&=\undersl{C}_{1}\mathrm{e}
^{-uh}
\label{Stetigkeit 4}
\end{align}

\noindent
Since the magnetic field between circular loop and conductive layer is of
interest, the quantity $\undersl{B}_{2}$ has to be determined. Eliminating of
$\undersl{A}_{1}$, $\undersl{A}_{2}$ and $\undersl{C}_{1}$ yields:

\begin{equation}
\dfrac{\undersl{B}_{1}}{\undersl{B}_{2}}=\dfrac{\left[\mu+\mu_{0}(1+\mathrm{j}k
)\right]\sinh(uh\sqrt{1+\mathrm{j}k})}{\left[\mu-\mu_{0}(1+\mathrm{j}k)\right
]\sinh(uh\sqrt{1+\mathrm{j}k})}
+\dfrac{2\mu\mu_{0}\sqrt{1+\mathrm{j}k}\cosh(uh\sqrt{1+\mathrm{j}k})}{\left[\mu
-\mu_{0}(1+\mathrm{j}k)\right]\sinh(uh\sqrt{1+\mathrm{j}k})}
\label{B2 zu B1}
\end{equation}

\noindent
The axial component of the magnetic field within the plane of the circular loop
$\undersl{H}_{z}^{\prime}(r,z=-d,t)=\undersl{H}_{\text{coil}}^{\prime}$ in
presence of the conductive layer yields with Eq.~(\ref{Biot23})

\begin{equation}
\undersl{H}_{\text{coil}}^{\prime}=\mathrm{e}^{\;\!\mathrm{j}\omega
t}\int\limits_{0}^{\infty}J_{0}(ur)\left[\undersl{B}_{1}\mathrm{e}^{ud}
+\undersl{B}_{2}\mathrm{e}^{-ud}\right] \, \mathrm{d}u \text{,}
\label{Biot25}
\end{equation}

\noindent
where $\undersl{B}_{1}$ and $\undersl{B}_{2}$ are defined by Eq.~(\ref{Biot20})
and
Eq.~(\ref{B2 zu B1}), respectively.

\subsection{Relative Inductance Change of a Circular Loop in Presence of a
Conductive Layer}
\label{ssec:05}

On the basis of the definition for the magnetic flux within the plane of the
circular loop in accordance with

\begin{equation}
\undersl{\varPhi}_{\text{coil}}=2\uppi\mu_{0}\int\limits_{0}^{a}\undersl{H}_
{\text{coil}}r\,\mathrm{d}r \text{,}
\label{Fluss 1}
\end{equation}

\noindent
a conductive layer influences the magnetic field of the coil
$\undersl{H}_{\text{coil}}$ resulting in a change of the magnetic flux:

\begin{equation}
\Delta\!\;\undersl{\varPhi}=\undersl{\varPhi}_{\text{coil}}-\undersl{\varPhi}_
{\text{coil}}^{\prime}=2\uppi\mu_{0}\int\limits_{0}^{a}(\undersl{H}_{\text{coil
}}-\undersl{H}_{\text{coil}}^{\prime})r\,\mathrm{d}r
\label{Fluss 2}
\end{equation}

\noindent
Insertion of Eqs.~(\ref{Biot22}) and (\ref{Biot25}) into Eq.~(\ref{Fluss 2})
yields
the
change of the magnetic flux at position $z=-d$

\begin{equation}
\Delta\!\;\undersl{\varPhi}=-2\uppi\mu_{0}\mathrm{e}^{\;\!\mathrm{j}\omega
t}\int\limits_{0}^{\infty}\int\limits_{0}^{a}J_{0}(ur)\undersl{B}_{2}\mathrm{e}
^{-ud}r\,\mathrm{d}r \,\mathrm{d}u
\label{Fluss 3}
\end{equation}

\noindent
and thus

\begin{equation}
\Delta\!\;\undersl{\varPhi}=-2\uppi\mu_{0}\mathrm{e}^{\;\!\mathrm{j}\omega
t}\int\limits_{0}^{\infty}\int\limits_{0}^{a}\dfrac{\undersl{B}_{2}}{\undersl{B
}_{1}}J_{0}(ur)\undersl{B}_{1}\mathrm{e}^{-ud}r\,\mathrm{d}r \,\mathrm{d}u
\text{.}
\label{Fluss 4}
\end{equation}

\noindent
Insertion of definition (\ref{Biot20}) for $\undersl{B}_{1}$ into
Eq.~(\ref{Fluss 4}) results in

\begin{equation}
\Delta\!\;\undersl{\varPhi}=-\uppi\mu_{0}a\undersl{I}\mathrm{e}^{\;\!\mathrm{j
}\omega t}\int\limits_{0}^{\infty}\int\limits_{0}^{a}
\dfrac{\undersl{B}_{2}}{\undersl{B}_{1}}J_{1}(ua)\mathrm{e}^{-2ud}urJ_{0}(ur)\
,\mathrm{d}r \,\mathrm{d}u \text{.}
\label{Fluss 5}
\end{equation}

\newpage
\noindent
The evaluation of the inner integral results in~\cite{Physiker}:

\begin{equation}
\int\limits_{0}^{a}urJ_{0}(ur)\,\mathrm{d}r=aJ_{1}(ua)
\label{Fluss 6}
\end{equation}

\noindent
Thus, Eq.~(\ref{Fluss 5}) can be written in following form:

\begin{equation}
\Delta\!\;\undersl{\varPhi}=-\uppi\mu_{0}a^2\undersl{I}\mathrm{e}^{\;\!\mathrm
{j}\omega t}\int\limits_{0}^{\infty}
\dfrac{\undersl{B}_{2}}{\undersl{B}_{1}}\mathrm{e}^{-2ud}J_{1}^2(ua)\,\mathrm{d
}u
\label{Fluss 7}
\end{equation}

\noindent
Due to the change of the magnetic flux, a voltage $\Delta\!\;\undersl{U}$ is
induced resulting in

\begin{equation}
\Delta\!\;\undersl{U}=\dfrac{\mathrm{d}\Delta\varPhi}{\mathrm{d}t}=\mathrm{j
}\omega\Delta\!\;\undersl{\varPhi}\text{.}
\label{Voltage 1}
\end{equation}

\noindent
Insertion of Eq.~(\ref{Fluss 7}) into Eq.~(\ref{Voltage 1}) yields the
impedance
change
$\Delta\!\;\undersl{Z}$:

\begin{equation}
\dfrac{\Delta\!\;\undersl{U}(t)}{\undersl{I}\mathrm{e}^{\;\!\mathrm{j}\omega
t}}=\Delta\!\;\undersl{Z}=\Delta R+\mathrm{j}\omega\Delta L
=-\mathrm{j}\omega\uppi\mu_{0}a^2\int\limits_{0}^{\infty}
\dfrac{\undersl{B}_{2}}{\undersl{B}_{1}}\mathrm{e}^{-2ud}J_{1}^2(ua)\,\mathrm{d
}u
\label{Voltage 2}
\end{equation}

The inductance of an idealized, free circular loop with small wire radius
$r_{\text{w}}$ compared to the loop radius $a$ can be represented according
to~\cite{Impedance}:

\begin{equation}
L_{0}=\dfrac{\uppi a\mu_{0}}{2}
\label{L0}
\end{equation}

\noindent
Thus, from Eq.~(\ref{Voltage 2}) follows

\begin{equation}
\dfrac{\Delta L}{L_{0}}+\dfrac{\Delta R}{\mathrm{j}\omega
L_{0}}=-2a\int\limits_{0}^{\infty}
\dfrac{\undersl{B}_{2}}{\undersl{B}_{1}}\mathrm{e}^{-2ud}J_{1}^2(ua)\,\mathrm{d
}u \text{.}
\label{DeltaL}
\end{equation}

\noindent
Introducing the dimensionless variable

\begin{equation}
z^{*}=2ud\text{,}
\label{z}
\end{equation}

\noindent
(\ref{DeltaL}) can be written as

\begin{equation}
\dfrac{\Delta L}{L_{0}}+\dfrac{\Delta R}{\mathrm{j}\omega
L_{0}}=-\dfrac{a}{d}\int\limits_{0}^{\infty}
\dfrac{\undersl{B}_{2}}{\undersl{B}_{1}}\mathrm{e}^{-z^{*}}J_{1}^2(\dfrac{az^{*
}}{2d})\,\mathrm{d}z^{*} \text{.}
\label{DeltaL2}
\end{equation}

\noindent
Comparing the left side of Eq.~(\ref{DeltaL2}) with the right side gives
evidence,
that the quotient of both integration constants $\undersl{B}_{1}$ and
$\undersl{B}_{2}$ has to be formulated as a complex number

\begin{equation}
\dfrac{\undersl{B}_{1}}{\undersl{B}_{2}}=\alpha(z^{*})+\mathrm{j}\beta(z^{*})
\text{.}
\label{complex}
\end{equation}

\noindent
In order to calculate the real part $\alpha(z^{*})$ and the imaginary part
$\beta(z^{*})$, a representation of the complex root $\sqrt{1+\mathrm{j}k}$ in
Eq.~(\ref{B2 zu B1}) is needed as a complex number by means of the
following
relationship~\cite{Loss}:

\begin{equation}
2\sqrt{1+\mathrm{j}k}=\lambda+\mathrm{j}\nu
\label{complex2}
\end{equation}

\noindent
Since the representation of the quotient in Eq.~(\ref{B2 zu B1}) as a
complex
number according to (\ref{complex}) comes along with an enormous mathematical
calculation complexity, the software package Mathematica was used, in order to
ease calculation significantly. Furthermore, the relative inductance change in
Eq.~(\ref{DeltaL2}) is of interest, thus a representation of $\beta(z^{*})$ in
Eq.~(\ref{complex}) is set aside.

\newpage
In combination with Eq.~(\ref{B2 zu B1}), Eq.~(\ref{complex}) and
Eq.~(\ref{complex2}),
the
fundamentals for the calculation of the relative inductance change of a
circular loop in presence of a conductive layer are finally achieved:

\begin{equation}
\dfrac{\Delta L}{L_{0}}=-\dfrac{a}{d}\int\limits_{0}^{\infty}
\mathrm{e}^{-z^{*}}J_{1}^2(\dfrac{az^{*}}{2d})\alpha(z^{*})\,\mathrm{d}z^{*}
\text{,}
\label{DeltaL3}
\end{equation}

\noindent
where

\begin{equation}
\alpha(z^{*})=-\dfrac{k}{D}\left[k\cosh\left(\dfrac{\lambda h
z^{*}}{2d}\right)+\nu\sinh\left(\dfrac{\lambda h z^{*}}{2d}\right)\right]
+\dfrac{k}{D}\left[k\cos\left(\dfrac{\nu h
z^{*}}{2d}\right)+\lambda\sin\left(\dfrac{\nu h z^{*}}{2d}\right)\right]
\label{alpha}
\end{equation}

\noindent
with

\begin{align}
D=\lambda^2\left(1+\dfrac{\lambda^2}{4}\right)\cosh\left(\dfrac{\lambda h
z^{*}}{2d}\right)+\lambda^3\sinh\left(\dfrac{\lambda h z^{*}}{2d}\right)
+\nu^2\left(1-\dfrac{\nu^2}{4}\right)\cos\left(\dfrac{\nu h
z^{*}}{2d}\right)+\nu^3\sin\left(\dfrac{\nu h z^{*}}{2d}\right)\text{,}
\label{De}
\end{align}

\begin{equation}
\dfrac{\lambda^2}{2}=\sqrt{1+k^2}+1\text{,}\hspace{10mm}\dfrac{\nu^2}{2}=\sqrt
{1+k^2}-1
\label{Def}
\end{equation}

\noindent
and
\begin{equation}
k=\dfrac{8\uppi d^2 f\mu_{0} \kappa}{(z^{*})^2}\text{.}
\label{Def2}
\end{equation}

\noindent
It should be noted, that Eq.~(\ref{DeltaL3}) is not solvable in closed form
analysis, thus a numeric integration has to be considered.

\section{Measurement Setups and Results}
\label{sec:05}

\subsection{Relative Inductance Change vs. Tip Deflection}
\label{ssec:01}

The relative inductance change of the sensing coil in presence of a conductive
\ce{Cu}-layer applied to the bender's tip is experimentally determined as a
function
of the distance $z$. The measurement results are compared with the results of
the analytical calculations based on the fundamental Eq.~(\ref{DeltaL3}).\\

\begin{figure}[h]
\centering
\includegraphics[scale=1]{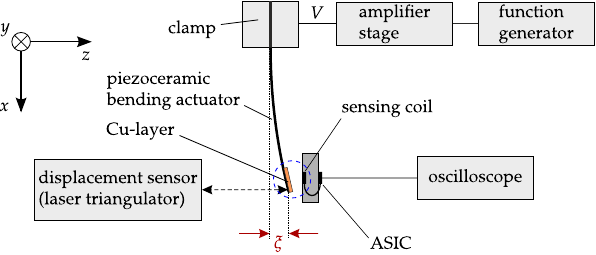}
\caption{Measurement setup for the experimental
determination of the relative inductance change of the sensing coil dependent
on
the tip deflection $\xi$ of a piezoceramic bending actuator}
\label{fig:Inductive7}
\end{figure}

The determination of the relative inductance change of the sensing coil with
respect to the conductive \ce{Cu}-layer is realized with the measurement setup
shown in Fig.~\ref{fig:Inductive7}.

\begin{figure}[h]
\centering
\includegraphics[scale=1]{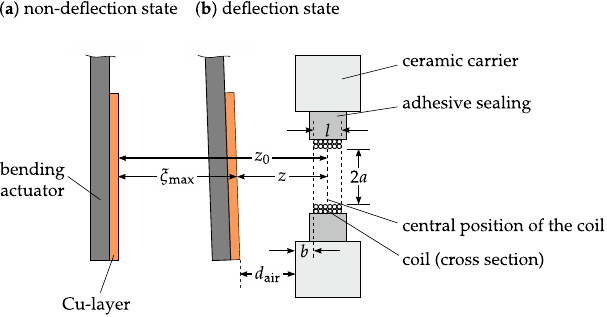}
\caption{Geometrical dimensions of the
sensor-actuator-system: (\textbf{a}) Non-deflection state. (\textbf{b})
Deflection state of the piezoceramic bending actuator.}
\label{fig:Inductive8}
\end{figure}

\noindent
Since the geometry of the arrangement plays a fundamental role concerning the
analytical calculations, the circle sector sketched in
Fig.~\ref{fig:Inductive7} is represented in higher magnification in
Fig.~\ref{fig:Inductive8} showing the non-deflection and deflection state of
the
piezoelectric bending actuator.

\noindent
Here, $z_{0}$ denotes the maximum distance between the upper side of the
\ce{Cu}-layer and the central position of the sensing coil for the
non-deflection
state of the bending actuator, $z$ denotes the distance between the upper side
of the \ce{Cu}-layer and the central position of the sensing coil for the
deflection
state of the bending actuator, $\xi_{\text{max}}$ corresponds to the maximum
deflection of the bending actuator, $b$ corresponds to the distance between the
front edges of the coil and the ceramic carrier, $a$ denotes the coil radius,
$l$ corresponds to the coil length and $d_{\text{air}}$ denotes the air gap
between ceramic carrier and \ce{Cu}-layer at maximum deflection.
The physical dimensions of the quantities $\xi_{\text{max}}$, $b$, $a$, $l$ and
$d_{\text{air}}$ being necessary for the analytical calculations of the
relative inductance change of the sensing coil in presence of the \ce{Cu}-layer
are
listed in Table~\ref{table_Quantities}.

\begin{table}[h]
\renewcommand{\arraystretch}{1.5}
\newcolumntype{P}[1]{>{\centering\arraybackslash}p{#1}}
\caption{Relevant Geometrical Quantities }
\label{table_Quantities}
\centering
\begin{tabular}{|P{1.6cm}|P{1.6cm}|P{1.6cm}|P{1.6cm}|P{1.6cm}|}
	\hline
$\xi_{\text{max}}$\,/\,\si{\micro\metre} 	& $b$\,/\,\si{\micro\metre}	&
$a$\,/\,\si{\micro\metre}		& $l$\,/\,\si{\micro\metre}	 	&
$d_{\text{air}}$\,/\,\si{\micro\metre} \\
	\hline
	329.6 										& 341.2						& 513.0							& 600.0							& 15.0 								   	\\
	\hline
\end{tabular}
\end{table}

\noindent
Taking Fig.~\ref{fig:Inductive8} into consideration, the following
correlation can be formulated for the deflection state of the bending actuator:

\begin{equation}
\xi_{\text{max}}+d_{\text{air}}+b+\dfrac{l}{2}=z_{0}
\label{Auslenkung}
\end{equation}

\noindent
Furthermore, for an arbitrary deflection $0<\xi<\xi_{\text{max}}$, it can be
written

\begin{equation}
\xi+z=z_{0} \text{,}
\label{Auslenkung 1}
\end{equation}

\noindent
thus, the distance $z$ between \ce{Cu}-layer and central position of the
sensing
coil can be calculated by insertion of Eq.~(\ref{Auslenkung 1}) into
Eq.~(\ref{Auslenkung}):

\begin{equation}
z=\xi_{\text{max}}+d_{\text{air}}+b+\dfrac{l}{2}-\xi
\label{Auslenkung 2}
\end{equation}

\noindent
Taking the Eqs.~(\ref{DeltaL3})-(\ref{Def2}) into consideration, the relative
inductance
change is analytically calculated. The variable $d$ in the
Eqs.~(\ref{DeltaL3})-(\ref{De}) and (\ref{Def2}) has to be replaced by the
variable
$z$ in accordance with Eq.~(\ref{Auslenkung 2}).
\\

The starting point for the experimental characterization of the relative
inductance change is represented by following equation defined already in
Eq.~(\ref{rel}):

\begin{equation}
\frac{\Delta L_{1}}{L_{1}}=1-\left(\frac{f_{1}}{f_{\text{e}}(z)}\right)^2
\label{rel 2}
\end{equation}

\begin{figure}[h]
\centering
\includegraphics[scale=1]{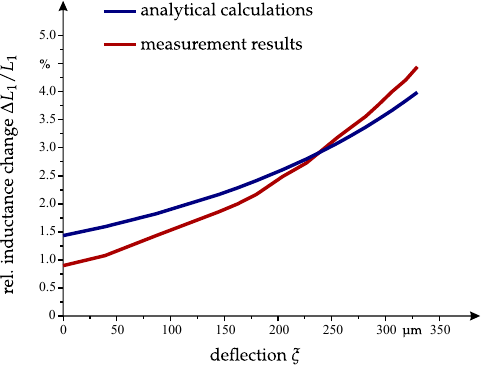}
\caption{Relative inductance change of the sensing coil as
a function of the actuator's tip deflection $\xi$.}
\label{fig:Inductive9}
\end{figure}

\noindent
Equation~(\ref{rel 2}) shows, that both, the knowledge of frequency $f_{1}$ of
the uninfluenced coil $(z\rightarrow\infty)$ and frequency $f_{\text{e}}(z)$ of
the coil affected by the \ce{Cu}-layer in distance $z$ is necessary. The
frequency
measurements concerning $f_{1}$ and $f_{1}(z)$ are achieved with an
oscilloscope. Since the distance $z$ is directly correlated with the deflection
$\xi$ of the bending actuator with respect 
to Eq.~(\ref{Auslenkung 2}), the
relative inductance change of the sensing coil can be represented as a function
of the deflection $\xi$ of the bending actuator. Figure~\ref{fig:Inductive9}
shows the experimentally determined and analytically calculated relative
inductance change of the
sensor-actuator-system as a function of the tip deflection $\xi$.

\subsection{Tip Deflection vs. Applied Driving Voltage}
\label{ssec:02}

\begin{figure}[h]
\centering
\includegraphics[scale=1]{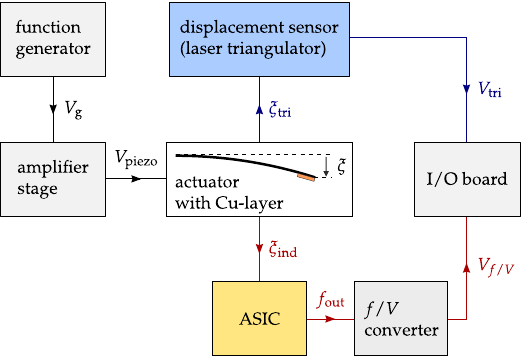}
\caption{Measurement setup for tip deflection measurement
of a piezoceramic bending actuator by means of an inductive proximity sensor.}
\label{fig:Inductive10}
\end{figure}

Since it is also an aim to detect the hysteresis of the piezoelectric bending
actuator by means of the integrated proximity sensor, the tip deflection has
been measured in accordance with the measurement setup shown in
Fig.~\ref{fig:Inductive10}.

\noindent
The amplifier stage following the function generator applies a sinusoidal
voltage $V_\text{piezo}$ with the frequency $f=\SI{200}{\milli\hertz}$ to the
piezoceramic bending actuator resulting in a bending movement of the cantilever
beam. Thus, the distance between the copper layer and the sensing coil
decreases and increases and the deflection $\xi$ of the bending actuator is
transformed to a frequency signal $f_\text{out}$ by the ASIC. In a next step,
the frequency signal is converted to a voltage signal $V_{f/V}$ by a
frequency-to-voltage converter and afterwards read in by an I/O board. The
deflection measurement always verified by a high precision laser deflection
measurement system provides a comparison and bench mark for the inductive
proximity sensor in combination with the frequency-to-voltage
converter~\cite{Eddy}.

\begin{figure}[h]
\centering
\includegraphics[scale=1]{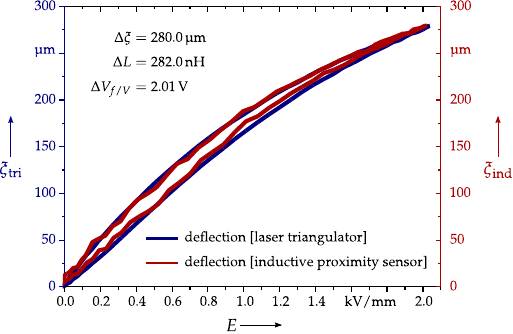}
\caption{Beam deflection $\xi$ and inductive sensor signal
$\xi_{\text{ind}}$ showing the deflection hysteresis versus applied electric
field $E$}
\label{fig:Inductive11}
\end{figure}

\noindent
Figure~\ref{fig:Inductive11} shows the tip deflection $\xi$ versus the electric
field $E$ (and thus the applied voltage $V_{\text{piezo}}$). The measurement
results show the ability of the inductive proximity sensor in combination with
the appropriate sensor electronics to detect the hysteresis being typical for
piezoceramic bending actuators.

\section{Deflection Measurement for Energy Harvesting Purposes}
\label{sec:06}

\begin{figure}[h]
\centering
\includegraphics[scale=1]{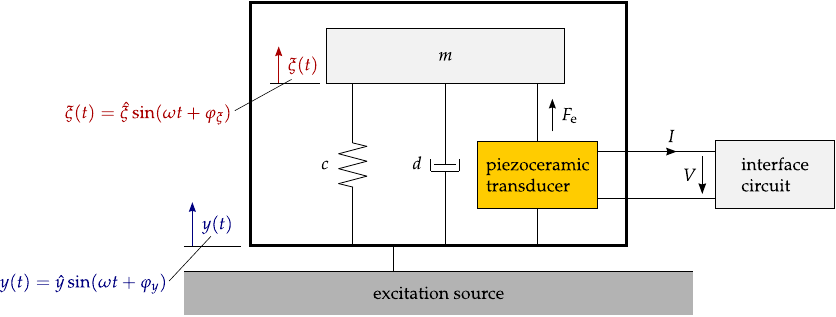}
\caption{Simplified model of a kinetic energy harvester~\cite{CMOS}.}
\label{fig:Inductive12}
\end{figure}

Measurement of the deflection $\xi$ of a cantilever bending actuator allows to
derive the amount of power dissipation for energy harvesting applications.
Piezoceramic bending actuators can be used as kinetic energy harvesters and are
modeled as second-order spring mass damper system in a simplified approach, see
Fig.~\ref{fig:Inductive12}.

Further simplification is necessary to derive the governing equation for
displacement of the amplitude of the system: it is assumed, that the mass of
the excitation source is much larger than seismic mass $m$ and that the
excitation source is practically not damped by the energy harvester. Thus, the
excitation source is assumed to provide infinite power. An external vibration
which is assumed to exert a sinusoidal force on the harvester frame, causes the
frame to move harmonically with

\begin{equation}
y(t)=\hat{y}\sin(\omega t+\varphi_{y})\text{,}
\label{y}
\end{equation}

\noindent
where $\hat{y}$ indicates the amplitude of the frame motion, and 
$\omega$
denotes the angular vibration frequency. The relative movement of the seismic
mass $m$ with respect to the frame is given by

\begin{equation}
\xi(t)=\hat{\xi}\sin(\omega t+\varphi_{\xi})\text{,}
\label{xi}
\end{equation}

\noindent
where $\xi$ indicates the amplitude of the mass motion and thus the bender's
tip deflection to be measured with the inductive proximity sensor. Since both,
the frame motion and the motion of mass represent harmonic quantities,
the Eqs.~(\ref{y}) and (\ref{xi}) can be expressed by complex amplitudes
according
to

\begin{equation}
\undersl{y}=\hat{y}\!\;\mathrm{e}^{\!\;\mathrm{j}\varphi_{y}}
\label{y_komplex}
\end{equation}
\noindent
and
\begin{equation}
\undersl{\xi}=\hat{\xi}\!\;\mathrm{e}^{\!\;\mathrm{j}\varphi_{\xi}}\text{.}
\label{xi_komplex}
\end{equation}

\begin{figure}[h]
\centering
\includegraphics[scale=1]{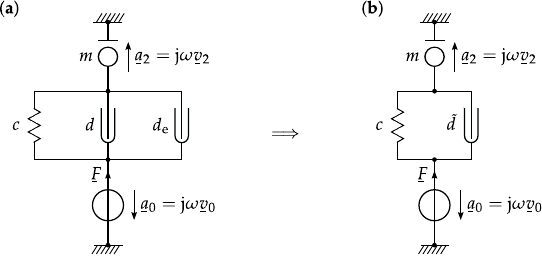}
\caption{Mechanical scheme of a kinetic energy harvester:
(\textbf{a}) With additional electrical damping $d_{\mathrm{e}}$ caused by the
connected interface circuit. (\textbf{b}) Equivalent damping 
$\tilde{d}$ as sum
of the mechanical damping $d$ and the electrical damping $d_{\mathrm{e}}$.}
\label{fig:Inductive13}
\end{figure}

Using complex amplitudes, the model of the kinetic energy harvester shown in
Fig.~\ref{fig:Inductive12} can be transformed into a mechanical scheme, see
Fig.~\ref{fig:Inductive13}a. Due to electromechanical feedback, the
transducer exerts a restoring force $F_{\mathrm{e}}$ on the seismic mass if an
interface circuit is connected, causing additional electrical damping
$d_\mathrm{e}$. The sum of the mechanical damping $d$ and the electrical
damping $d_{\mathrm{e}}$ is defined by the quantity $\tilde{d}$, see
Fig.~\ref{fig:Inductive13}b.

\begin{figure}[h]
\centering
\includegraphics[scale=1]{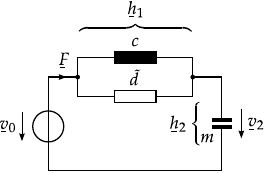}
\caption{Equivalent circuit representation of a kinetic
energy harvester.}
\label{fig:Inductive14}
\end{figure}

In a next step, the mechanical scheme shown in Fig.~\ref{fig:Inductive13}b is
transferred to the equivalent circuit representation according to
Fig.~\ref{fig:Inductive14}. Here a mechanical force is represented as an
electric
current, and a mechanical velocity can be replaced as an electrical voltage.
The mass is represented as a capacitor with the value $m$, the spring is
modelled as an inductor with the value $c$ (spring compliance) and the sum of
the mechanical and the electrical dampings is described by the resistor
$\tilde{d}$. The voltage source models the external vibration.

\noindent
Taking the mechanical admittances $\undersl{h}_{1}$ and 
$\undersl{h}_{2}$ in
Fig.~\ref{fig:Inductive14} into account, the ratio between the bender's tip
deflection $\undersl{\xi}$ and the movement $\undersl{y}$ of the frame results
in:

\begin{equation}
\dfrac{\undersl{v}_{2}}{\undersl{v}_{0}}=\dfrac{\mathrm{j}\omega\undersl{\xi}}
{\mathrm{j}\omega\undersl{y}}=\dfrac{\undersl{h}_{2}}{\undersl{h}_{1}+\undersl
{h}_{2}}=\dfrac{1+\mathrm{j}\frac{\omega}{\omega_{0}}\frac{1}{Q}}{1-\left(\frac
{\omega}{\omega_{0}}\right)^{2}+\mathrm{j}\frac{\omega}{\omega_{0}}\frac{1}{Q}}
\label{Geschwindigkeiten}
\end{equation}

\noindent
with the resonance frequency

\begin{equation}
\omega_{0}=\dfrac{1}{\sqrt{mc}}
\label{Resonanzfrequenz}
\end{equation}

\noindent
and the quality factor
\begin{equation}
Q=\dfrac{1}{\omega_{0}c\tilde{d}}\,\text{.}
\label{Q}
\end{equation}

\noindent
The displacement amplitude of the seismic mass $m$ can be written as:

\begin{equation}
\hat{\xi}(\omega)=\dfrac{\sqrt{1+\left(\frac{\omega}{\omega_{0}}\frac{1}{Q
}\right)^{2}}}{\sqrt{\left[1-\left(\frac{\omega}{\omega_{0}}\right)^{2}\right]^
{2}+\left(\frac{\omega}{\omega_{0}}\frac{1}{Q}\right)^{2}}}\,\hat{y}(\omega)
\label{eq:mdisplacement}
\end{equation}

\noindent
Thus, the amplitude $\hat{y}$ of the frame motion can be determined by
measuring the bender's tip deflection $\hat{\xi}$.
\\

\begin{figure}[h]
\centering
\includegraphics[scale=1]{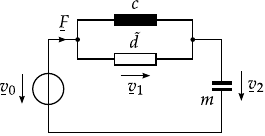}
\caption[Inductive sensor 12]{Determination of the power dissipated in the
transducer.}
\label{fig:Inductive15}
\end{figure}

In order to determine the power dissipated in the transducer, i.e. in the
resistor $d_{\mathrm{e}}$, Fig.\ref{fig:Inductive15} is considered.

\noindent
It can be written:
\begin{equation}
P=\dfrac{1}{2}\undersl{F}_{1}\undersl{v}_{1}^{*}
=\dfrac{1}{2}d_{\mathrm{e}}\undersl{v}_{1}\undersl{v}_{1}^{*}=\dfrac{1}{2}d_
{\mathrm{e}}\left|\undersl{v}_{1}\right|^{2}
\label{P}
\end{equation}

\noindent
The velocity $\undersl{v}_{1}$ is gained by the the velocities of the frame an
the bender's tip and thus by the amplitudes of the frame motion and the tip
deflection:

\begin{equation}
\undersl{v}_{1}=\undersl{v}_{0}-\undersl{v}_{2}=\mathrm{j}\omega\left(\undersl
{y}-\undersl{\xi}\right)
\label{P2}
\end{equation}

\noindent
Since the amplitude of the frame motion can be determined by means of measuring
the tip deflection according to Eq.~(\ref{eq:mdisplacement}), also the quantity
$\undersl{v}_{1}$ and thus the power dissipated in the transducer according to
Eq.~(\ref{P}) can be determined.

\section{Conclusion}
\label{sec:07}
In this paper, we focused on the electrodynamics of a non-contact inductive
proximity sensor for high-accuracy tip deflection measurements of piezoelectric
bending actuators. The sensor principle itself is based on eddy current
effects, offering the possibility of sensor integration in smallest spaces in
combination with a high-accuracy electronic circuit. In order to determine the
relative change of the inductance as a function of the beam bender's tip
deflection in an analytical way, an electrodynamic approach was formulated
verifying the experimental measurement results. Besides the ability to detect
the hysteresis, which represents a typical characteristic of piezoceramic
bending actuators, the measurement of the displacement via a non-contact
inductive sensor system has also the potential to directly evaluate the
possible amount of energy which can be harvested from a piezoelectric energy
harvester. In future, optimization of material and mechanical shape may be
investigated and instantly evaluated in means of performance by measurement of
the deflection and the external equivalent electrical parameters.

\reftitle{References}

\end{document}